\documentclass[]{spie}  

\usepackage{amsmath,amsfonts,amssymb}
\usepackage{graphicx}
\usepackage[colorlinks=true, allcolors=blue]{hyperref}
\usepackage{multirow}
\usepackage{orcidlink}
\newcommand{\nelec}{n_{\rm e}}

\title{Radiation damage to the Hubble Space Telescope\\ has been several years out of phase with the Solar cycle}

\author[a]{Gavin Leroy\,\orcidlink{0009-0004-2523-4425}}
\author[b]{Juan Paolo Lorenzo Gerardo Barrios\,\orcidlink{0000-0002-5605-0029}}
\author[a]{\\Maximilian von Wietersheim-Kramsta\,\orcidlink{0000-0003-4986-5091}}
\author[a]{Richard Massey\,\orcidlink{0000-0002-6085-3780}}
\author[a]{\\Richard G.\ Hayes}
\author[c]{Jacob A.\ Kegerreis\,\orcidlink{0000-0001-5383-236X}}
\author[d]{David Lagattuta\,\orcidlink{0000-0002-7633-2883}}
\author[a]{Zane D.\ Lentz\,\orcidlink{0000-0003-4428-7843}}
\author[e]{\\James W.\ Nightingale\,\orcidlink{0000-0002-8987-7401}}
\author[f]{Jesper Skottfelt\,\orcidlink{0000-0003-1310-8283}}
\author[g]{Felix Vecchi\,\orcidlink{0009-0004-7808-1979}}
\affil[a]{\small{Institute for Computational Cosmology, Durham University, South Road, Durham DH1 3LE, UK}}
\affil[b]{Cavendish Laboratory, University of Cambridge, JJ Thomson Avenue, Cambridge CB3 0HE, UK}
\affil[c]{Department of Earth Science and Engineering, Imperial College London, London SW7 2BP, UK}
\affil[d]{Centre for Astrophysics Research, Department of Physics, University of Hertfordshire, Hatfield AL10 9AB, UK}
\affil[e]{Physics Department, Newcastle University, Newcastle upon Tyne NE1 7RU, UK}
\affil[f]{Centre for Electronic Imaging, The Open University, Walton Hall, Milton Keynes MK7 6AA, UK}
\affil[g]{Laboratoire d’Astrophysique, EPFL, Observatoire de Sauverny, 1290 Versoix, Switzerland \normalsize}

\authorinfo{Send correspondence to J.P.L.G.B. at \href{mailto:jplgmb2@cam.ac.uk}{jplgmb2@cam.ac.uk}.} 

\begin{document} 
\maketitle

\begin{abstract}
As well as obtaining beautiful images of the Universe, the {\sl Hubble Space Telescope}'s CCD detectors are sensitive radiation dosimeters that have been monitored in Low Earth Orbit for more than 24~years. The rate of radiation damage they received has varied over each Solar cycle, but several years out of phase with the appearance of sunspots or coronal mass ejections. We investigate functional forms that successfully fit the time series of damage to telescopes elsewhere in the Solar system. 
We obtain remarkably accurate fits to {\sl Hubble} data but with physically absurd parameter values. During image post-processing, such fits can be used empirically, to correct more than 99.5\% of the radiation damage's effect on image quality. However, fits to the time series with physically reasonable parameters produce worse performance. 
Our results highlight the diversity of radiation environments in different parts of our Solar system, and the complexity of Low Earth Orbit in particular. Our results also motivate continued monitoring of radiation damage to currently operational spacecraft, to more reliably predict the rate of degradation in (and useful lifespan of) future missions.
\end{abstract}


\section{INTRODUCTION}
\label{sec:intro}  

\begin{figure} [t]
   \begin{center}
   \begin{tabular}{cc} 
      \raisebox{1.6cm}{
         \includegraphics[width=6.2cm]{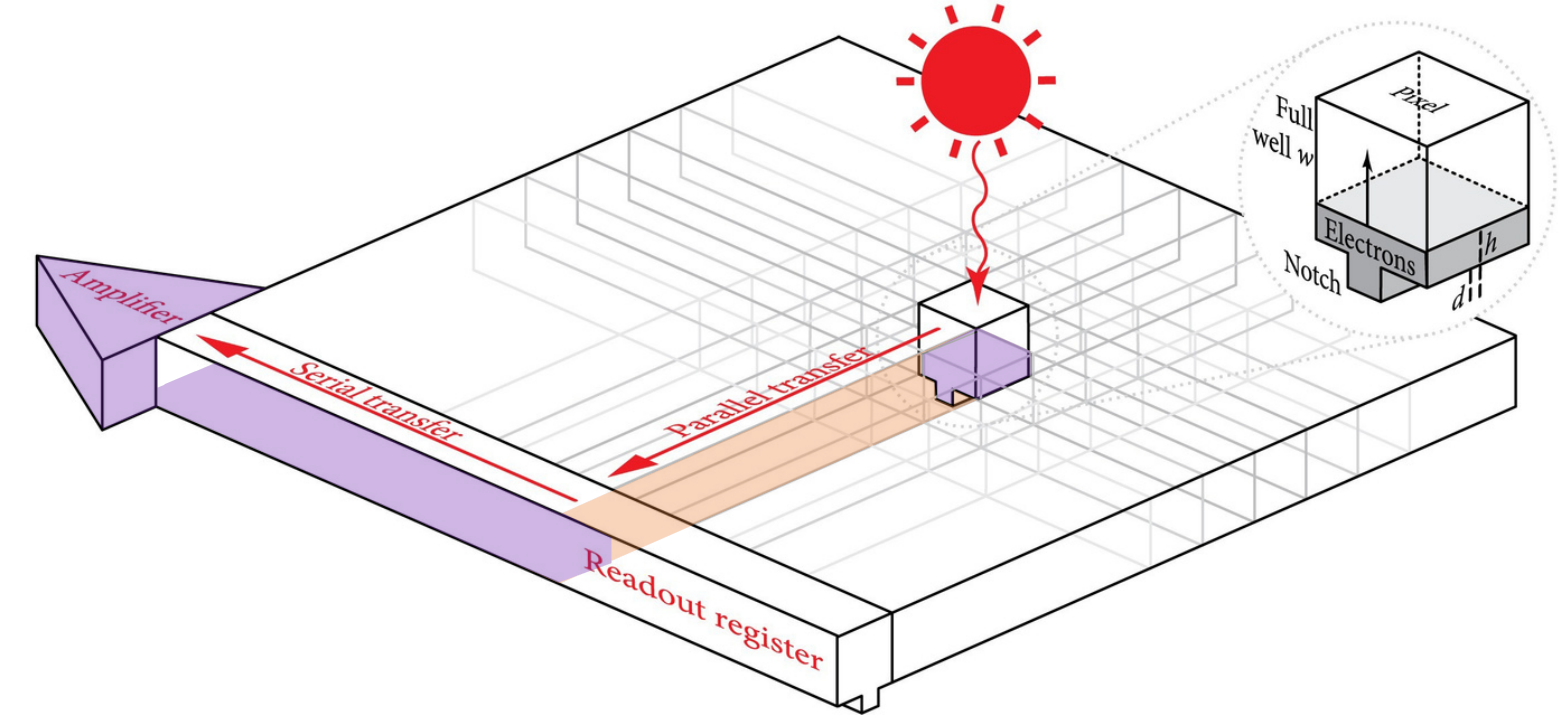}
      } &
      \includegraphics[width=10cm]{CTI_trailing_1000DPI_squarezoom_ADU_compressed.png}
   \end{tabular}
   \end{center}
   \caption[CCD trailing]
   { \label{fig:trailing} 
{\it Left:} Photoelectrons created in a CCD pixel are counted at the end of an exposure by shifting them in the parallel then serial direction to an amplifier and ADC at the corner. {\it Right:} Radiation damage to the {\sl Hubble Space Telescope} has created defects that delay electron flow for a duration similar to the pixel-to-pixel transfer time in the parallel direction\cite{Anderson2024,Ryon2024}. This creates spurious trailing behind all image features whose electrons have to move past many traps.}
\end{figure}

Above the protection of the Earth's atmosphere, Charge-Coupled Device (CCD) imaging detectors are gradually damaged by the harsh radiation environment. High-energy charged particles displace atoms from the silicon wafer, creating lattice defects that disrupt the smooth transport of photoelectrons during detector readout \cite{Holland1990}. The defects temporarily capture electrons and release them after characteristic delays \cite{ShockleyRead1952,Hall1951}, shifting charge away from its original location and producing spurious trails behind astronomical sources \cite{Janesick1987,Jerram2020,Ali2022} as illustrated in figure~\ref{fig:trailing}. Several species of defect can be created, corresponding to different topological configurations of dislocated atoms. Each species of trap delays charge for a different time, superimposing trails of different length. 
This spurious trailing, which depends non-linearly on source brightness, morphology, and illumination history \cite{Rhodes2007}, can be the most serious obstacle to some scientific measurements \cite{Cropper2013,Soto2023,Astier2023,Stark2024,2025hst..prop17925B,plato2025}.

\begin{figure} [t]
   \begin{center}
   \begin{tabular}{l} 
      \includegraphics[width=0.9\textwidth]{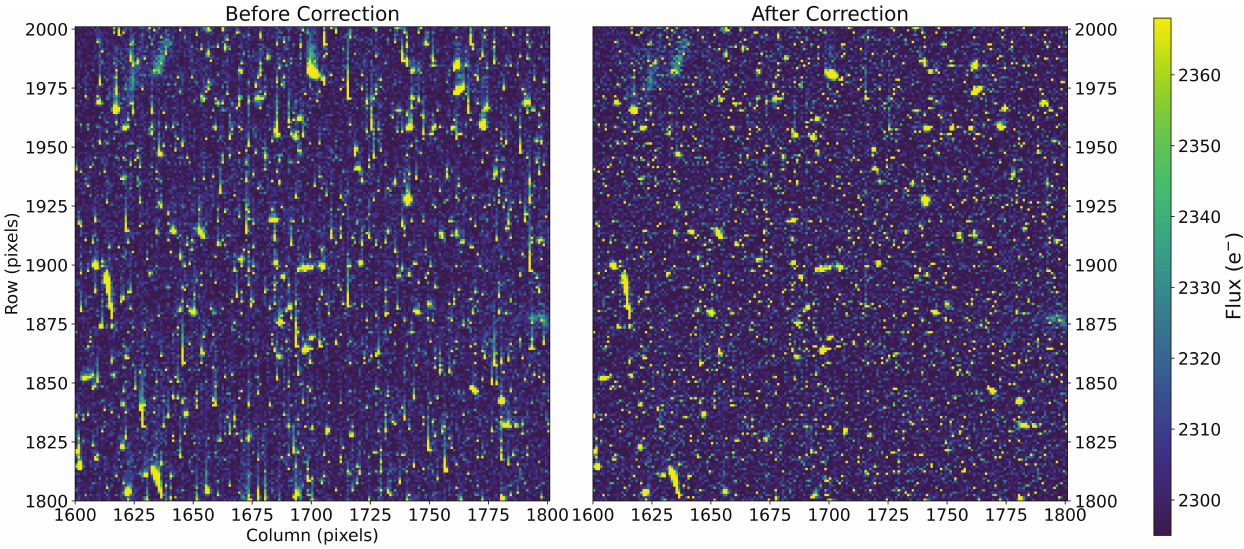}
   \end{tabular}
   \end{center}
   \caption[Corrected CCD trailing]
   { \label{fig:correction} 
{\it Left:} An image acquired by the {\sl Hubble Space Telescope} in January 2025, before any processing. Note the spurious trailing above every image feature, caused by radiation-induced Charge Transfer Inefficiency. 
{\it Right:} The same image, corrected following the procedure in Massey et al.\ (2026).}
\end{figure}

The physics of radiation damage to to silicon is understood well. 
Models of electron flow through damaged CCDs can reproduce the observed trailing, and these models can be inverted to restore the true image\cite{Bristow2003}. 
Following continuous study over its 24 years in space, gradually improving models of the {\sl Hubble Space Telescope} ({\sl HST}) {\sl Advanced Camera for Surveys/Wide Field Channel} ({\sl ACS/WFC}), \cite{Massey+2010,Massey2010,Anderson2010,Massey+2014,Chiaberge2022} now correct better than 99.5\% of image trailing\cite{Massey+2026} (figure~\ref{fig:correction}).

{\sl Hubble}'s CCD detectors are thus (amongst other uses) phenomenally precise radiation dosimeters that have been continuously monitored in Low Earth Orbit for longer than two Solar cycles (figure~\ref{fig:time_series}). Measurements of the gradual degradation of CCD performance in various astronomical telescopes may inform models of the radiation environment at different locations within our Solar system, such as SPENVIS \cite{heynderickx2000spenvis,messios2025spenvis}.

In this paper, we assess a new explanation for the curious time series of CCD degradation in {\sl Hubble} (Section~\ref{sec:time_series}). We also describe recent improvements to the model of electron flow, which have stabilised its performance in extreme regimes, improving the correction of bias and dark frames as well as science exposures (Section~\ref{sec:arctic}). We conclude in Section~\ref{sec:conclusions}.

\section{TIME EVOLUTION OF RADIATION DAMAGE}
\label{sec:time_series}

We study a time series of the mean density of charge traps per {\sl ACS/WFC} pixel, $\rho_{\mathrm{trap}}(t)$ that cause CTI, as measured by Massey et al.\ (2026)\cite{Massey+2026}. The timing and the relative amplitude of damage measured in this way roughly matches measurements of damage from the growth rate of sink pixels in the same CCDs\cite{Guzman2024}.

\begin{figure} [t]
   \begin{center}
   \begin{tabular}{l} 
      \includegraphics[width=\textwidth]{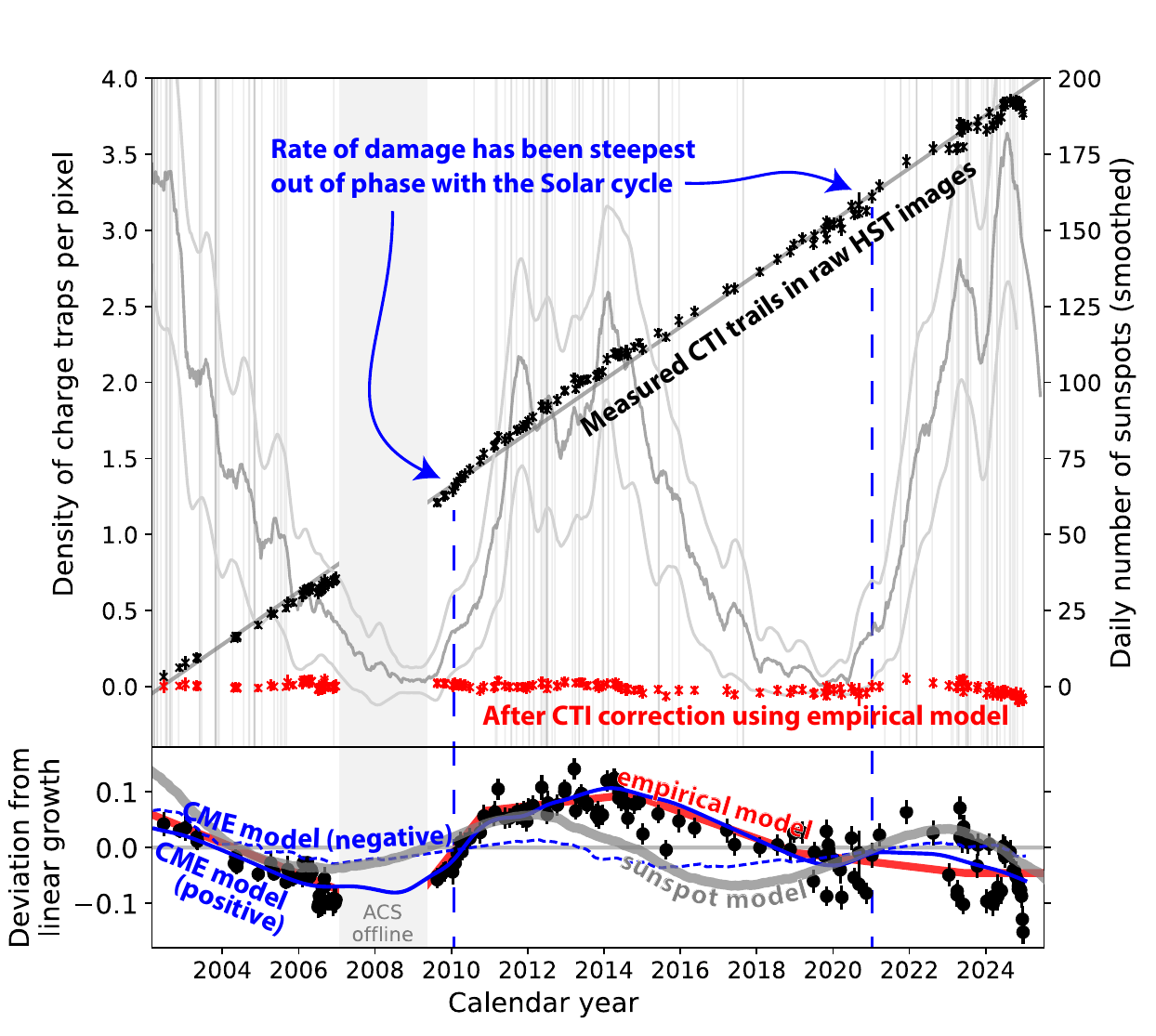}
   \end{tabular}
   \end{center}
   \caption[CCD trailing]
   { \label{fig:time_series} 
{\it Top:} Measurements of the accumulated damage to CCD detectors onboard the Hubble Space Telescope (black points), from the amplitude of trails behind warm pixels\cite{Massey+2026}. The grey band shows the mean number of sunspots in a 3~month moving window (centered on the date shown), and the standard deviation of the number of suspots per day within that window. Vertical grey lines indicate coronal mass ejection (CME) events with a flux of high energy particles $>$$10$~protons~cm$^{-2}$~sr$^{-1}$\,s$^{-1}$.
{\it Bottom:} The deviation of the damage from linear growth (black points), plus three models attempting to predict it. These include a model based on sunspots, with a 430~day lag (grey), a better-fitting model based on CMEs but with an unphysical 8 year lag (blue), and a pragmatic piecewise-linear empirical model (red) that has no physical motivation but enables 99\% of the imaging trailing to be corrected.}
\end{figure}

Curiously, the rate of degradation of {\sl Hubble}'s performance has been out of phase with the Solar cycle. The rate of damage varies by $\sim$18.5\% over each 11~year period, but the maximum rate of damage occurs approximately 4.3~years before Solar maximum. Massey et al.\ (2026) fit a model using daily numbers of sunspots, $n_\mathrm{sunspot}(t)$ from the SILSO World Data Center\cite{sidc},
\begin{equation}
    \rho_{\mathrm{trap}}(t) = \rho_0 + A_\mathrm{GCR} \times t + A_\mathrm{sunspot} \int_{t_\mathrm{0}}^t \Big( n_{\mathrm{sunspot}}(t'-t_\mathrm{lag}) \Big)^\eta~dt',
    \label{eqn:rhotot_sunspot}
\end{equation}
where $t$ is the time in days since launch at $t_0$. Best-fit values of parameters $\rho_0$,  $A_\mathrm{GCR}$, $A_\mathrm{sunspot}$, $\eta$ and $t_\mathrm{lag}$ are listed in table~\ref{tab:best_fit_params} and the fit is shown as a grey curve in figure~\ref{fig:time_series}: it is almost indistinguishable from the best-fit sinusoid. 
The negative sign of best-fit parameter $A_\mathrm{sunspot}$ implies that the appearance of sunspots {\it reduces} the rate of CCD degradation in Low Earth Orbit, as if the increased particle flux or Solar wind suppresses Galactic Cosmic Rays\cite{Kilifarska2020,Koldobskiy2022,Tahtinen2024}. Those authors were unable to distinguish between the $t_\mathrm{lag}=430^{+11}_{-5}$~day time lag being a phase difference in Solar physics between the production of energetic particles and sunspots, or a local delay caused by the Earth's magnetic field\cite{Hands2018,Matthia2023}. 

Here we try fitting a model in which the rate of growth of charge traps, $d\rho_{\mathrm{trap}}(t)/dt$, is assumed to correlate with Coronal Mass Ejection (CME) events. 
We use measurements of the fluence of particles with energy $>$10~MeV during CMEs, obtained by the NOAA {\sl Geostationary Operational Environmental Satellites} ({\sl GOES})\footnote{GOES data is from \href{https://www.ngdc.noaa.gov/stp/space-weather/interplanetary-data/solar-proton-events/SEP\%20page\%20code.html}{www.ngdc.noaa.gov/stp/space-weather/interplanetary-data/solar-proton-events}.} in Earth Geosynchronous orbit. 
Specifically, we fit 
\begin{eqnarray}
    \frac{d\rho_{\mathrm{trap}}}{dt}(t) = A_\mathrm{GCR} + \delta(t - t_{\mathrm{CME},i}) \times A_\mathrm{CME} \times (P_{\mathrm{CME},i}
    )
    ^\eta \\
\text{i.e.}~~~~
    \rho_{\mathrm{trap}}(t) = \rho_0 + A_\mathrm{GCR} \times t +  A_\mathrm{CME} \sum_i (P_{\mathrm{CME},i}
    )
    ^\eta\,,\,~~~~~~
\end{eqnarray}
where $P_\mathrm{CME}$ is the integral 5-minute average proton flux, in units of protons~cm$^{-2}$\,sr$^{-1}$\,s$^{-1}$, during an event at time $t_{\mathrm{CME},i}$ and the index $i$ runs over all such events since launch.
This model fits the degradation of CCDs in {\sl Euclid}\cite{Skottfelt2024} and {\sl Gaia}\cite{gaia26}\footnote{{\sl Gaia} data prefer small modulation of $A_\mathrm{GCR}$ over an period longer than the 10~year mission during (Claudio Pagani, priv.\ comm.); it would be very interesting to measure the period and phase of that with respect to the Solar cycle.} both of which are at Lagrange point L2, but gives a {\it terrible} fit to the measured values of $\rho_{\mathrm{trap}}(t)$ in {\sl Hubble} for any values of parameters $\rho_0$,  $A_\mathrm{GCR}$, $A_\mathrm{CME}$ and $\eta$. 

Qualitatively, a remarkably accurate fit can be obtained by allowing a delay $t_\mathrm{lag}$ between the CME and the damage (and softening the step functions to keep $\rho_{\mathrm{trap}}$ differentiable), via
\begin{equation}
    \rho_{\mathrm{trap}}(t) = \rho_0 + A_\mathrm{GCR} \times t + A_\mathrm{CME} \sum_{t_\mathrm{launch}}^t \frac{(P_{\mathrm{CME},i})
    ^\eta}{1+\exp{((t - t_{\mathrm{CME},i}-t_\mathrm{lag})/\lambda)}}~.
    \label{eqn:rhotot_cme}
\end{equation}
This function (solid blue curve in figure~\ref{fig:time_series}) even incorporates a previously-unexplained reduction in the rate of damage in 2025. 
However, the best-fit parameters (table~\ref{tab:best_fit_params}) include an 8~year lag between CME protons reaching Geosynchronous orbit and damage happening to detectors in Low Earth Orbit that is so long it must clearly be unphysical. 
This model's fit is nonetheless even more remarkable when compared to a fit using the pattern of sunspots in the subsequent Solar cycle. 
Forcing the fit to a local minimum near $t_\mathrm{lag}\approx 8-11=-3$~years fails to reproduce features in $\rho_{\mathrm{trap}}(t)$ (and is even more unphysical: the as-yet uncounted number of sunspots in 2027 are needed to model the accumulation of damage in 2024).

The best-fit value of $A_\mathrm{CME}$ is positive. 
If it is instead forced to be negative, as if the particle flux from the Sun during CME events suppresses Galactic Cosmic Rays reaching Low Earth Orbit, the phase of the damage is recovered with a $t_\mathrm{lag}=424.7\pm2.1$~day lag almost identical to that of the sunspot model (table~\ref{tab:best_fit_params}).
However, the poor overall fit (dotted blue curve in figure~\ref{fig:time_series}) fails to reproduce the dynamic range in the rate of damage. 
Even with a best-fit value of $\eta\approx 0$ (counting all CMEs equally effectively reproduces the sunspot numbers with coarser time resolution), this model cannot simultaneously match the steep excess rate of damage during $\sim$$2010$ and the shallow reduction in the rate of damage from $\sim$$2014$--$2020$.

\begin{table}[t]
\caption{Best-fit parameters for models of the time series of damage to {\sl Hubble}'s CCD detectors, $\rho_{\mathrm{trap}}(t)$. The sunspot model \eqref{eqn:rhotot_sunspot} produces the grey curve in figure~\ref{fig:time_series}, with parameter values fitted by Massey et al.\ (2026). The Coronal Mass Ejection model \eqref{eqn:rhotot_cme} with a postive/negative value of $A_\mathrm{CME}$ produces the solid/dotted blue curve in figure~\ref{fig:time_series}.} 
\label{tab:best_fit_params}
\begin{center}  
\begin{tabular}{|l|r@{ $\pm$ }l|} 
\hline
\rule[-1ex]{0pt}{3.5ex}  Parameter & \multicolumn{2}{c|}{Sunspot model~~~~~~~~~~}  \\
\hline
\rule[-1ex]{0pt}{3.5ex}  $\rho_0$ & $0.0898$ & $0.0045$  \\
\hline
\rule[-1ex]{0pt}{3.5ex}  $A_\mathrm{GCR}$ & $(5.5738$ & $0.0098)\times10^{-4}$  \\
\hline
\rule[-1ex]{0pt}{3.5ex}  $A_\mathrm{sunspot}$ & $(-3.6802$ & $0.0334)\times10^{-6}$ \\
\hline
\rule[-1ex]{0pt}{3.5ex}  $\eta$ & $0.8309$ & $0.0020$  \\
\hline 
\rule[-1ex]{0pt}{3.5ex}  $t_\mathrm{lag}$ & \multicolumn{2}{c|}{$430~^{+11}_{~-5}$~days~~~~~~} \\
\hline 
\multicolumn{3}{c}{~} \\
\end{tabular}
~~
\begin{tabular}{|l|r@{ $\pm$ }l|r@{ $\pm$ }l|} 
\hline
\rule[-1ex]{0pt}{3.5ex}  Parameter & \multicolumn{2}{c|}{CME model (positive)} & \multicolumn{2}{c|}{CME model (negative)} \\
\hline
\rule[-1ex]{0pt}{3.5ex}  $\rho_0$ & $-0.5785$ & $0.0014$ & $0.4644$ & $0.0018$ \\
\hline
\rule[-1ex]{0pt}{3.5ex}  $A_\mathrm{GCR}$ & $0.000390$ & $0.000002$ & $0.000499$ & $0.000001$ \\
\hline
\rule[-1ex]{0pt}{3.5ex}  $A_\mathrm{CME}$ & $0.000689$ & $0.000010$ & $-0.002465$ & $0.000033$ \\
\hline
\rule[-1ex]{0pt}{3.5ex}  $\eta$ & $0.3353$ & $0.0019$ & $0.00004$ & $0.00260$ \\
\hline 
\rule[-1ex]{0pt}{3.5ex}  $t_\mathrm{lag}$ & $3074$ & $1$ days & $424.7$ & $2.1$ days \\
\hline 
\rule[-1ex]{0pt}{3.5ex}  $\lambda$ & $89.2$ & $1.2$ days & $10.7$ & $1.1$ days \\
\hline 
\end{tabular}
\end{center}
\end{table}

This analysis leaves us with more questions than answers. Good fits are possible with apparently unphysical parameters; reasonable parameters and reasonable functional forms lead to qualitatively poor fits. Pragmatically, accurate correction for CTI is possible with any well-fitting model, including even a simple piecewise-linear fit (red curve and corrected data points in figure~\ref{fig:time_series}), but this approach is not useful to predict future performance.

\vspace{10mm}
\section{IMPROVEMENTS TO A MODEL OF ELECTRON TRANSPORT \\THROUGH A DAMAGED CCD}
\label{sec:arctic}

Simulating the passage of a cloud of $n_\mathrm{e}$ electrons through a CCD substrate to readout electronics requires three ingredients. These are a model of the volume (or cross-sectional area) of that cloud, $V(n_\mathrm{e})$, which determines how many traps it encounters, plus probabilities of capture by and release from a charge trap, which may be functions of time.
Here we report recent improvements in two of those ingredients.

\subsection{Cross-sectional volume of a cloud of electrons}
\label{sec:volume}

A simple parameterisation of the fraction of a pixel filled by a cloud of $\nelec$ electrons has often\cite{Rhodes+2010,Massey+2014,Massey+2026} been
\begin{equation}
  V(\nelec) =
  \begin{cases}
    \left( \dfrac{\nelec - d}{w - d} \right)^\beta & \text{for } \nelec > d \\[8pt]
    0 & \text{for } \nelec < d,
  \end{cases}
\end{equation}
where $w$ is the full depth, $d\geqslant0$ is the depth of a supplementary buried channel or `notch' built in to some detectors, and the power $\beta$ is typically $\sim$$0.5$.
However, this function is not differentiable at $\nelec=d$, which creates a very large cross-section for trailing of the next electron. The sudden change in gradient also creates asymmetry in trailing of e.g.\ bias or dark exposures when $\nelec\approx d$: positive noise fluctuations are exposed to traps, and are trailed away, but negative noise fluctuations are untrailed.

To rectify these asymmetries, we now allow $d<0$ and use
\begin{equation}
  V(\nelec) =
  \begin{cases}
    \left( \dfrac{\nelec - d}{w - d} \right)^\beta & \text{for } \nelec > 0 \text{ and } d<0 \\
    \left(\dfrac{-\,d}{w-d}\right)^\beta~\exp{\left( \dfrac{-\beta}{d}\nelec\right)} & \text{for } \nelec < 0 \text{ and } d<0,
  \end{cases}
  \label{eqn:volume_new}
\end{equation}
where the coefficient in the exponent ensures smoothness in $dV/d\nelec$ at $\nelec=0$.
The $\nelec=0$ location of the transition between these two curves is arbitrary, and could be another free parameter --- but empirically fitting it in the $\sim$zero signal regime would likely be impossible. Nonetheless, the now-finite gradient at $\nelec=0$ and the extension of the function to negative $\nelec$ (which is needed in the presence of measurement noise) suitably stabilises trailing in bias and dark exposures.  

\begin{figure} [t]
   \begin{center}
   \begin{tabular}{cc} 
      \includegraphics[width=8.2cm]{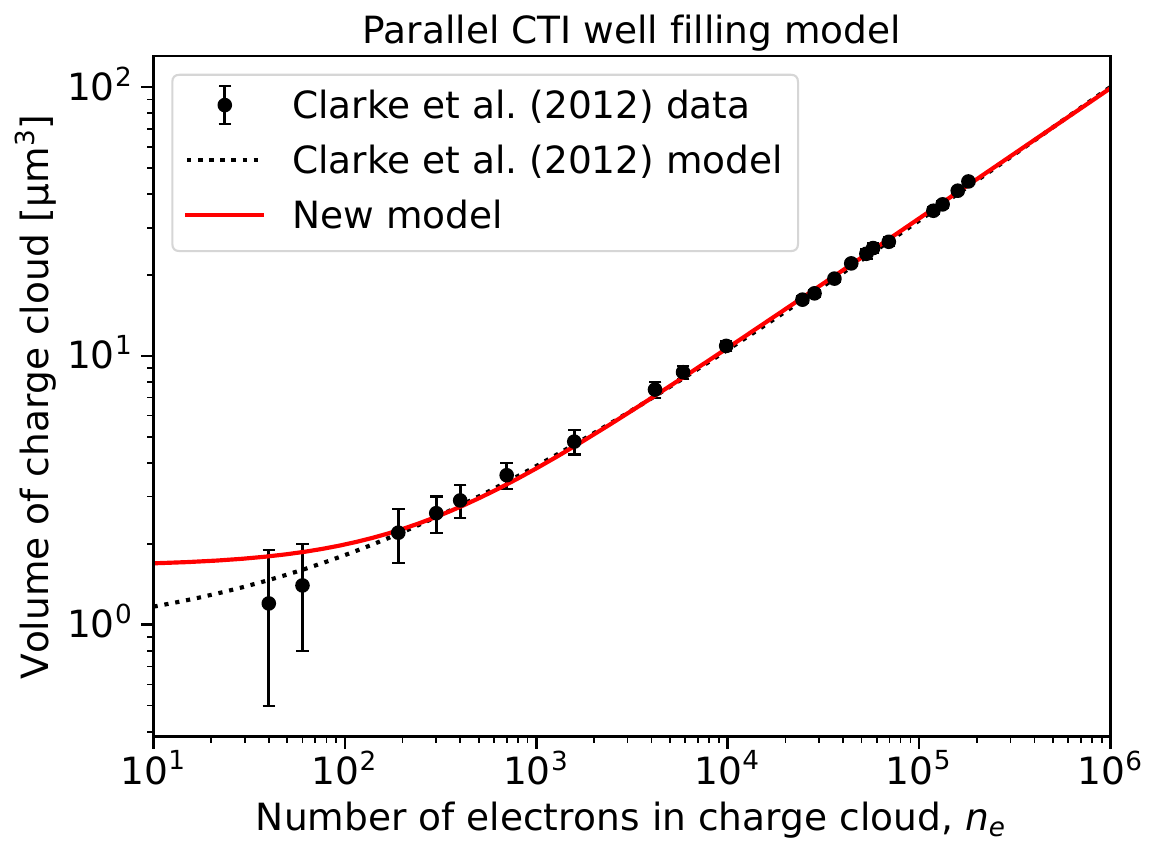}
      &
      \includegraphics[width=8.2cm]{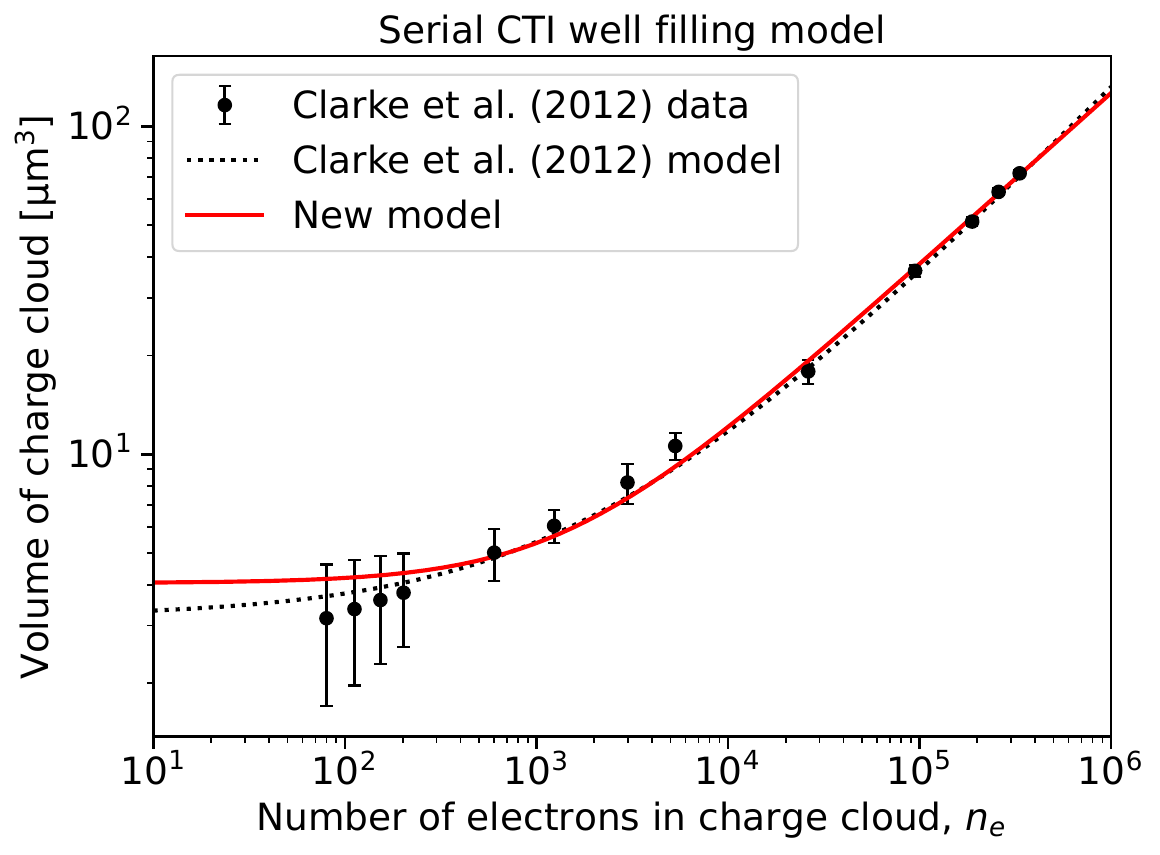}
   \end{tabular}
   \end{center}
   \caption[CCD trailing]
   { \label{fig:clarke} 
The effective volume of a cloud of $\nelec$ electrons, from more detailed simulations\cite{Clarke2025} of electron density within solid-state structure using Silvaco TCAD. Over the full range of signal levels, our new model (red curves, equation~\ref{eqn:volume_new} with parameters listed in table~\ref{tab:clarke}). is as good a fit as that proposed by Clarke et al.\ (dotted black curves). However, our new model usefully extends to negative $\nelec$, which is needed in the presence of measurement noise, and its finite gradient at $\nelec=0$ stabilises correction of CTI in bias or dark exposures where pixel values are close to zero.\vspace{5mm}}
\end{figure} 

In practice, we find that parameters $\beta$ and $d$ are highly degenerate when constrained using the trailing of warm pixels. Instead, we fit $\beta$ and $\upsilon\equiv\log_{10}{(\alpha)}\equiv\beta\log_{10}{(-d/(w-d))}$, which are approximately orthogonal.
The inspiration for this can be seen by rewriting the second case of equation~\eqref{eqn:volume_new} as\footnote{Using logarithms to base 10 rather than $e$ in the definition of $\upsilon$ is inelegant, but this choice is now locked in to the code.} $V(\nelec)=\exp{(\upsilon\ln{(10)}-\beta\nelec/d)}$.


\begin{table}[t]
\caption{Best-fit parameters of equation~\eqref{eqn:volume_new}, which models the effective volume of a cloud of $\nelec$ electrons. These produce the red curves in figure~\ref{fig:clarke}.} 
\label{tab:clarke}
\begin{center}       
\begin{tabular}{|l|r@{ $\pm$ }l|r@{ $\pm$ }l|} 
\hline
\rule[-1ex]{0pt}{3.5ex}  Parameter & \multicolumn{2}{c|}{Parallel} & \multicolumn{2}{c|}{Serial} \\
\hline
\rule[-1ex]{0pt}{3.5ex}  $\beta$ & $0.4855$ & $0.010$ & $0.5244$ & $0.023$ \\
\hline
\rule[-1ex]{0pt}{3.5ex}  $\upsilon=\log_{10}(\alpha)$ & $-1.439$ & $0.080$ & $-1.130$ & $0.058$  \\
\hline
\rule[-1ex]{0pt}{3.5ex}  ~~~~\dots inferred $d$ & \multicolumn{2}{c|}{$-218.1$} & \multicolumn{2}{c|}{$-1407.4$} \\
\hline 
\rule[-1ex]{0pt}{3.5ex}  $V(\nelec=w=200,000)$ & $45.4$ & $0.6\mu$m$^3$ & $54.6$ & $0.7\mu$m$^3$\\
\hline 
\end{tabular}
\end{center}
\end{table}

The form of volume-filling function \eqref{eqn:volume_new} is inspired by more detailed solid-state Silvaco TCAD\cite{silvaco} simulations of electron density within an (albeit different) CCD\cite{Clarke2025}, reproduced here in figure~\ref{fig:clarke}. Whereas we achieve algorithmic stability by shifting the base power law horizontally, they chose to shift it vertically, as $V=\gamma_c (\nelec)^{\beta_c} + \alpha_c$. They obtain best-fit values $\{\alpha_c=0.89$, $\beta_c=0.51$, $\gamma_c=0.09\}$ for parallel transport and $\{\alpha_c=3.195$, $\beta_c=0.59$, $\gamma_c=0.04\}$ for the serial register. Approximating
\begin{equation}
  \beta\approx\beta_c \text{~~~and~~~}
  d = \frac{w}{1 - \alpha^{(-1/\beta)}} 
   \approx \frac{w}{1 - (1+\gamma_c w^{\beta_c} /\alpha_c)^{1/\beta_c}}
   \approx -(\alpha_c/\gamma_c)^{(1/\beta_c)}  
\end{equation}
suggests we could use $d\approx-86$ for parallel, and $d\approx-1530$ for serial.
However, explicitly refitting their data (approximately recovered via \href{https://plotdigitizer.com}{plotdigitizer.com}) yields the parameters in table~\ref{tab:clarke} and red curves in figure~\ref{fig:clarke}. Both functional forms fit the data well, although the logarithmic axes emphasise their differences at the low end.

\subsection{Probability of electrons being released from charge traps}
\label{sec:continuum}

Charge traps typically release captured electrons with a characteristic half-life, producing an (approximately) exponential trail profile. The characteristic release time, $\tau$ typically takes one of several values, depending on the topological configuration of displaced silicon atoms (such as a vacancy, divacancy, or interstitial defect), or the atomic number of an impurity. However, if the CCD is kept cryogenically cold, trap pumping measurements in laboratories\cite{Gow2016,Bilgi2019,Parsons2021} and in space\cite{Skottfelt2024,gaia26} suggest newly-generated traps develop a broadened, roughly lognormal distribution of $tau$ around the preferred values. This is probably because the silicon atoms anneal only slowly towards the common, lowest-energy configurations. 

The lognormal distribution $N(\tau)$ can be explained as a normal distribution of trap energy band gaps $E$,
\begin{equation}
    N(E)\propto \exp{\dfrac{-(E-E_i)^2}{2\,\sigma_i^2}},
\end{equation}
where $E_i$ is the lowest energy state and $\sigma_i$ is some scatter reflecting the lack of annealing. Following solid-state theory\cite{ShockleyRead1952, Hall1951, Janesick1987}, a trap's characteristic release time is $\tau\propto e^{E/kT}$ at temperature $T$, so
\begin{eqnarray}
    N(\log\tau)~d\log\tau &=& N(E)~ dE \\
    N(\log\tau) &=& \dfrac{dE}{d\log\tau} N(E) 
    ~~=~~ \dfrac{kT}{\log{e}} \exp{\dfrac{-(E-E_i)^2}{2\,\sigma_i^2}} \nonumber \\
    ~ &=& \frac{kT}{\log{e}} 
    \exp{\left( \dfrac{-(\log\tau-\log\tau_i)^2}{2\left(\frac{\log{e}}{kT}~\sigma_i\right)^2} \right)} ~.
\end{eqnarray}
i.e.\ a lognormal distribution in $\tau$.
A distribution of release times perturbs the profile of charge trails. Nonetheless, we continue to find no statistically significant evidence for $\sigma_i>0$ in measurements of the shape of trails (in either {\sl Hubble} or {\sl Euclid}). 

\section{Conclusions}
\label{sec:conclusions}

Models of Charge Transfer Inefficiency in radiation-damaged CCDs continue to improve. 
Once calibrated using in-orbit data, they enable increasingly comprehensive correction during data postprocessing\cite{Massey+2026}. 
However, the rate of degradation over time remains unsatisfactorily unpredictable. 
Measurements from several spacecraft have now shown that the rate of degradation depends upon Solar activity --- but that it is not necessarily in phase with the appearance of sunspots or coronal mass ejections.
The qualitatively different functions of time that are required to model the degradation of CCDs in Low Earth Orbit or at Lagrange point L2 also highlights the diversity of radiation environments in different parts of our Solar system. 

To predict the future performance of missions like {\sl PLATO}\cite{plato2025}, and as STScI prepare for {\sl Hubble}'s continuing operation in to the 2030s\footnote{\href{https://docs.google.com/forms/d/e/1FAIpQLSe0ICn0kcRYjDE141dY0y1-O_9gqxewsLSAORMPCYFVStGfrQ/viewform}{https://hst-docs.stsci.edu/hstos}}, we recommend that telescope operation teams continue in-situ measurements of CTI, using all the capabilities of modern detectors such as electronic injection of charge into pre-defined patterns.
As well as being useful for all that astronomy and earth-imaging stuff, CCDs are highly sensitive radiation dosimeters that can teach us about the higher-energy radiation environment throughout our Solar system.


\acknowledgments 
 
Authors were supported in Durham by the UK STFC (grant ST/X001075/1) and the UK Space Agency (grant ST/X001997/1).
JAK and JWN are supported by STFC Ernest Rutherford Fellowships, and ZDL is supported by STFC studentship ST/Y509346/1.
Our statistical analysis software was partially developed through STFC grant ST/T002565/1, then  InnovateUK grants TS/V002856/1 and TS/Y014693/1.

\bibliography{cti, report} 
\bibliographystyle{spiebib} 

\end{document}